\documentclass[preprint,pteplogo]{ptephy_v2}

\preprintnumber{XXXX-XXXX} 

\makeatletter
\renewcommand\ps@headings{\let\@mkboth\markboth
\renewcommand\@oddhead{\hbox to \textwidth{\ifpteplogo\includegraphics[scale=0.40]{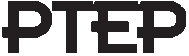}\else\fi\hfill}}
\renewcommand\@oddfoot{\hbox to \textwidth{\hfil\rmfamily\thepage\hfil%
\makebox[0\p@][r]{\@typeset}}}}
\makeatother

\usepackage{hyperref}

\usepackage{here}
\usepackage{url} 
\usepackage{subcaption}
\usepackage{lineno}

\begin{document}


\title{Search for Sub-GeV Axion-Like Particles at EBES Pilot Run Using 4~GeV Positron Beam at KEK LINAC}


\author{Takahiro Fusayasu}
\affil{Department of Physics, Saga University, Saga, Japan}

\author{Tomoya Iizawa}
\affil{International Center for Elementary Particle Physics and Department of Physics, The University of Tokyo, Tokyo, Japan}

\author[2]{Fumihito Ikeda}

\author{Akimasa Ishikawa}
\affil{High Energy Accelerator Research Organization (KEK), Tsukuba, Japan}

\author{Masako Iwasaki}
\affil{Osaka Metropolitan University, Osaka, Japan}

\author[3]{Hiroshi Iwase}

\author[2]{Takahiro Kawahara}

\author[1]{Aoi Masaki}

\author[3]{Fusashi Miyahara}

\author[3]{Yu Morikawa}

\author[3]{Yuichi Okayasu}

\author[1]{Toshiyuki Ono}

\author{Hidetoshi Otono}
\affil{Department of Physics, Kyushu University, Fukuoka, Japan}

\author[3]{Yasuhito Sakaki}

\author[2]{Takumi Seino}

\author[1]{Yuta Shimasaki}

\author[2]{Taikan Suehara}

\author{Yosuke Takubo}
\affil{National Institute of Technology, Niihama College, Niihama, Japan}

\author[5]{Shusaku Tsumura}

\author[4]{Kosuke Uemura}

\author[2]{Yifu Zhang}

\begin{abstract}
We report the results of a search for sub-GeV axion-like particles (ALPs) using pilot run data from the Electron Beam-dump Experiment at KEK LINAC Switching Yard 3 (EBES). The data were collected in December 2023 with a 4~GeV positron beam and correspond to \(5.7\times10^{13}\) positrons on target. In the pilot run setup, a tungsten beam dump and a single PbO calorimeter were used. We consider ALP production via the Primakoff process induced by bremsstrahlung photons in the beam dump, followed by the decay \(a\to\gamma\gamma\). The background was estimated with a data-driven method, and a signal region was defined such that the expected background yield is below 0.1 events. No events were observed after unblinding. Upper limits at the 90\% confidence level were derived in the ALP mass--coupling plane, extending the experimental coverage into a region of parameter space not explored by previous searches.
\end{abstract}

\subjectindex{C15, C31}

\maketitle

\section{Introduction}
The search for weakly coupled particles in the sub-GeV mass range has become increasingly important for exploring physics beyond the Standard Model (SM). Such particles may exist in a so-called ``dark sector", where they interact only weakly with SM particles. One prominent candidate in this category is Axion-Like Particles (ALPs), a hypothetical neutral pseudoscalar boson predicted by theories that extend the SM~\cite{Peccei:1977hh,Weinberg:1977ma,Wilczek:1977pj,Svrcek:2006yi,Arvanitaki:2009fg}. Since sub-GeV ALPs can serve as candidates for dark matter or as the inflaton responsible for cosmic inflation, they are the subject of extensive experimental searches~\cite{Beacham:2019nyx,Antel:2023hkf}.

EBES (Electron Beam-dump Experiment at KEK LINAC Switching Yard 3) \cite{Ishikawa:2021qna} is a new beam-dump experiment designed to search for axion-like particles (ALPs) using 4 GeV positron or 7 GeV electron beams produced at the KEK LINAC \cite{Akemoto:2013lxa}\cite{Furukawa:2022wzc}. ALPs are expected to be produced through interactions between the positron or electron beam and the target material. EBES is located at Switching Yard 3 (SY3) of the LINAC. The experimental setup consists of a tungsten beam dump equipped with a cooling system, a decay volume in which ALPs are expected to decay into two photons, and a detector system comprising a preshower detector with two silicon pixel layers and four PbO calorimeters. The preshower detector is used to identify photon pairs from ALP decays, while the PbO calorimeters measure their energies.

In the winter of 2023, we performed a pilot run at SY3 to measure background levels at the experimental site, using a tungsten beam dump and a single PbO calorimeter that are also employed in the full EBES setup. During this data-taking period, the operational parameters of the LINAC accelerator were optimized to minimize background levels. The collected data also contributed to the development of the analysis software. In addition, to demonstrate the capability of EBES to search for ALPs, an ALP search was performed using the pilot run data and the developed analysis software. This paper presents the results of the ALP search conducted with the pilot run data.

\section{Signal Model}
\label{sec:signal_model}

In this study, we consider pseudoscalar ALPs (\(a\)) coupled to photons, and adopt the following effective Lagrangian:
\begin{equation}
\mathcal{L}_{\rm int}\supset -\frac{1}{4}g_{a\gamma\gamma}\,a\,F_{\mu\nu}\tilde F^{\mu\nu}.
\label{eq:lagrangian_agammagamma}
\end{equation}
Here \(F_{\mu\nu}\) is the electromagnetic field-strength tensor, \(\tilde F^{\mu\nu}=\tfrac12\epsilon^{\mu\nu\lambda\rho}F_{\lambda\rho}\) is its dual, and \(g_{a\gamma\gamma}\) denotes the photon--ALP coupling constant. 
In what follows, the ALP mass ($m_a$) and the photon--ALP coupling constant ($g_{a\gamma\gamma}$) are treated as independent effective parameters.

As the dominant production mechanism in a beam-dump experiment, we consider ALP production via the Primakoff effect with a real photon in the initial state,
\begin{equation}
\gamma + N \to a + N.
\label{eq:primakoff_process}
\end{equation}
The initial-state photons are predominantly those produced as bremsstrahlung photons from electrons and positrons in the electromagnetic shower developed in the target by the primary beams.

Within this model, the ALP decay width is given by
\begin{equation}
\Gamma_{a\to\gamma\gamma}=\frac{g_{a\gamma\gamma}^2\,m_a^3}{64\pi}.
\label{eq:width_agammagamma}
\end{equation}
In the laboratory frame, the lifetime is dilated by the Lorentz factor \(\gamma\), and the characteristic decay length is expressed as
\(l_a = \beta\gamma /\Gamma_{a\to\gamma\gamma}\),
where natural units with \(\hbar = c = 1\) are used.

The kinematics of the Primakoff process is dominated by small momentum transfer \(t\), for which the nuclear elastic form factor \(F(t)\) plays an important role. For high-\(Z\) targets, we adopt a commonly used parameterization of \(F(t)\), following Refs.~\cite{Tsai:1986tx,Dusaev:2020gxi}, that incorporates both screening by atomic electrons and finite nuclear-size effects, and use the resulting differential cross section in the laboratory frame,
\begin{equation}
\frac{d\sigma}{dt}
=\frac{1}{2^3}\,g_{a\gamma\gamma}^2\,\alpha\,F^2(t)\,
\frac{t-t_{\min}}{t^2},
\qquad
t_{\min}=\frac{m_a^4}{4E_\gamma^2},
\label{eq:primakoff_dsigdt}
\end{equation}
where \(E_\gamma\) is the incident-photon energy and \(\alpha\) is the fine-structure constant.
As described in Sec.~\ref{chap:samples}, the above expressions are used in the signal simulation.

\section{Experimental setup}
\subsection{Overview}
\begin{figure}[H]
\centering
\includegraphics[width=0.62\linewidth]{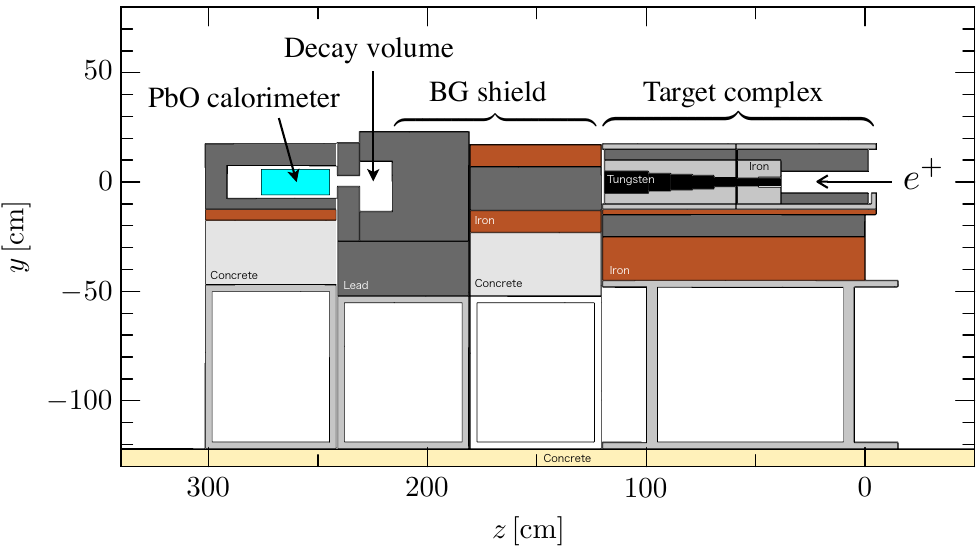}\hfill
\includegraphics[width=0.355\linewidth]{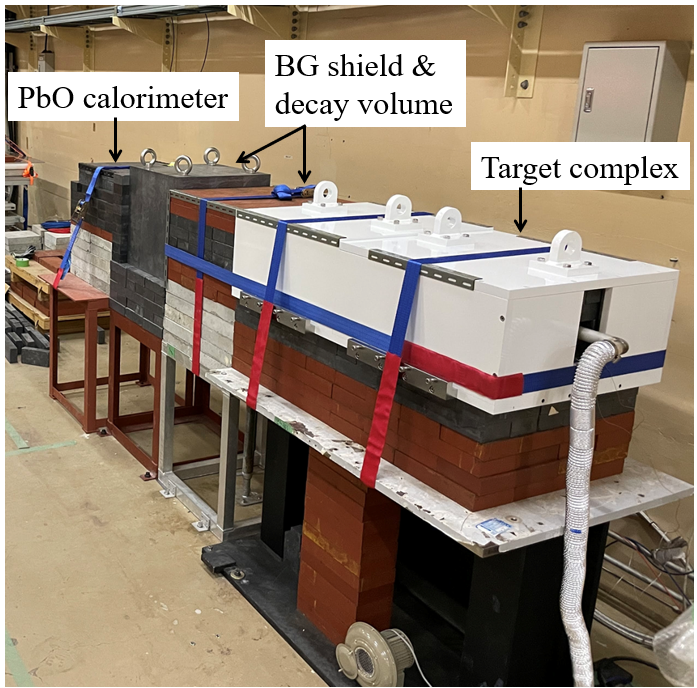}
\caption{The schematic view (left) and picture (right) of the pilot run setup of the EBES experiment.}
\label{fig:setup}
\end{figure}

In the pilot run, a 4~GeV positron beam produced at KEK LINAC is injected into a tungsten target to measure the background level at the EBES site and to demonstrate the capability to search for ALPs in the EBES experiment. The experimental setup for the pilot run consists of a target complex, a background shield, a decay volume with a 28 cm air gap, and a PbO calorimeter, arranged from upstream to downstream, as shown in Fig.~\ref{fig:setup}. Except for the background shield and the decay volume, all components are used in the final setup of the EBES experiment. In this section, each experimental component is described.

\subsection{LINAC}
\label{chap:LINAC}
The KEK e\textsuperscript{-}/e\textsuperscript{+} LINAC supplies electron beams to SuperKEKB HER (High Energy Ring), PF (Photon Factory), and PF-AR (PF Advanced Ring), and positron beams to SuperKEKB LER (Low Energy Ring). The electron beam can be accelerated up to 7~GeV, while the positron beam can reach energies of up to 4~GeV. The machine operates at a repetition rate of 50~Hz and supports arbitrary injection patterns according to the requirements of each storage ring. The LINAC can switch, on a pulse-by-pulse basis, among multiple beam modes according to the beam destination, such as downstream circular accelerators (SuperKEKB HER/LER, PF, and PF-AR) or the SY3 beam dump. In this experiment, we used a beam mode that transports the positron beam, produced via pair production in a tungsten target from a primary electron beam generated by a thermionic gun, to the SY3 dump. The repetition rate of this beam mode was up to 16.7 Hz, and the machine was operated with single bunch per pulse.

\begin{figure}[H]
\centering
\includegraphics[width=0.96\linewidth]{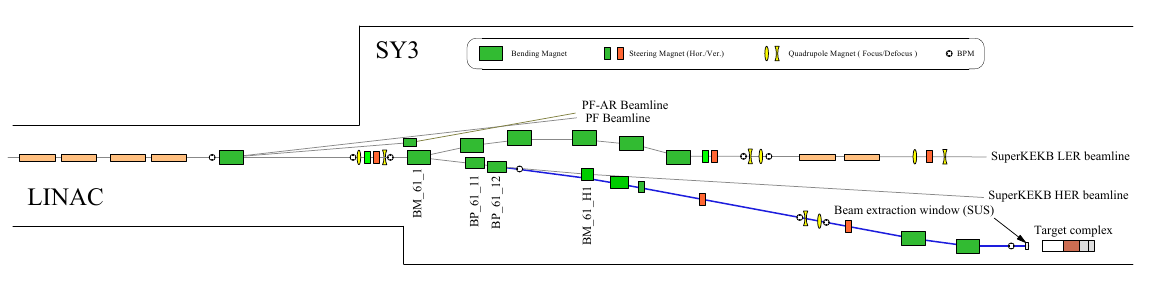}
\caption{Overview of the beamlines in SY3 and the location of the EBES target complex.}
\label{fig:SY3}
\end{figure}

The beams delivered to each ring are separated at SY3 into their respective beam transport lines. A dedicated beam diagnostic line is installed at SY3, and the target complex of this experiment is also used as the beam dump for the diagnostic line. An overview of SY3 is shown in Fig.~\ref{fig:SY3}. If beam loss occurs upstream of the target complex, high-energy secondary particles can enter the decay volume and the side surfaces of the PbO calorimeter, thereby contributing to the experimental background. Therefore, minimizing beam loss upstream of the target complex is essential. To monitor beam loss along the beamline, fiber-based loss monitors were installed from the bending magnet (BM\_61\_1), which transports the beam to the diagnostic line, to the beam extraction window.

When the beam tail or halo collides with the beam duct or other beamline components, a large number of secondary particles are produced. As these electrons pass through the loss monitor, Cherenkov radiation is generated. The Cherenkov light is guided to a photomultiplier tube (PMT) installed at the upstream end of the fiber, and the PMT output signals are recorded using an oscilloscope and an ADC module.

\begin{figure}[H]
\centering
\includegraphics[width=0.96\linewidth]{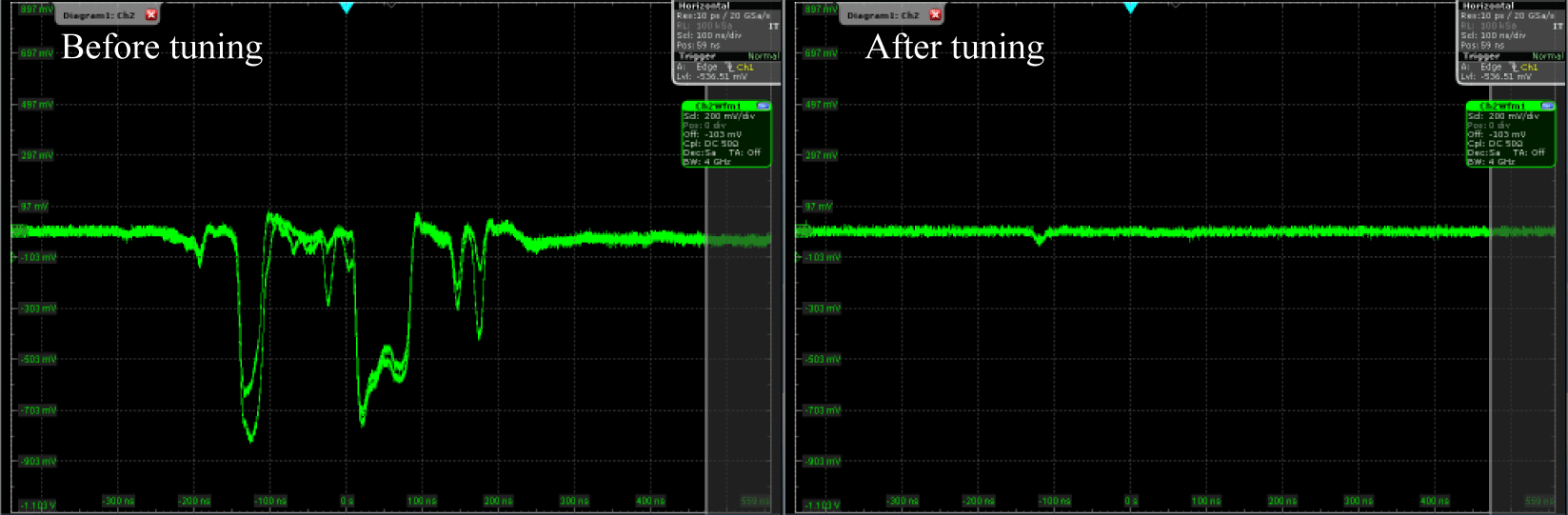}
\caption{Oscilloscope traces of the fiber loss monitor signals measured before beam tuning (left) and after beam tuning (right). The horizontal axis corresponds to the position of the fiber laid along the beamline. The persistence mode display shows the beam loss over several pulses.}
\label{fig:loss-monitor-signal}
\end{figure}

Figure~\ref{fig:loss-monitor-signal} shows the loss monitor signals before and after beam tuning. 
The loss monitor trigger is synchronized with the beam trigger, and the oscilloscope images are accumulated overlays over several tens of minutes or longer, corresponding to several thousand shots. Beam losses are mainly observed in the beam ducts immediately downstream of bending magnets with large dispersion, as well as at quadrupole magnets where the duct aperture is smaller. Before tuning, significant beam losses were observed at multiple locations. Because the beam orbit and beam size are stable, nearly the same amount of beam loss occurs at approximately the same location from shot to shot. These losses are considered to originate primarily from the tails of the beam distribution caused by the energy spread. By optimizing the RF phases of the upstream accelerating structures to control the energy spread, and by adjusting the beam orbit and beta function, these beam losses were successfully minimized.

\subsection{Target complex}
The primary beam is injected into a cylindrical tungsten target with a length of \(L=20~\mathrm{cm}\) and a diameter of \(\phi=3.5~\mathrm{cm}\). The beam was tuned so that its size at the target entrance was 1 mm or less. The radiation length and the Moli\`ere radius of tungsten are $X_0 = 0.35$~cm and $R_M = 0.93$~cm, respectively, and the target length and radius correspond to approximately $57\,X_0$ and $1.9\,R_M$. The target is surrounded by iron and lead shielding to suppress background events and to reduce secondary radiation leaking into the experimental area. The target is air-cooled, and its temperature is continuously monitored during operation. Immediately downstream of the target, a tungsten block with a length of \(60~\mathrm{cm}\) is installed, which is also enclosed by iron and lead shielding. Because the target region becomes strongly activated, the target assembly is modularized as much as possible and is designed to be removed quickly when needed for maintenance and replacement.

\subsection{Background shield and decay volume}
Downstream of the target complex, a \(95~\mathrm{cm}\)-long lead shield is placed along the beam axis to reduce beam-related backgrounds. Between the downstream end of this lead shield and the PbO calorimeter, there is a \(28~\mathrm{cm}\) air gap, which we define as the decay volume.

\subsection{PbO calorimeter} \label{sec:pbo}
\begin{figure}[H]
\centering
\includegraphics[width=10cm]{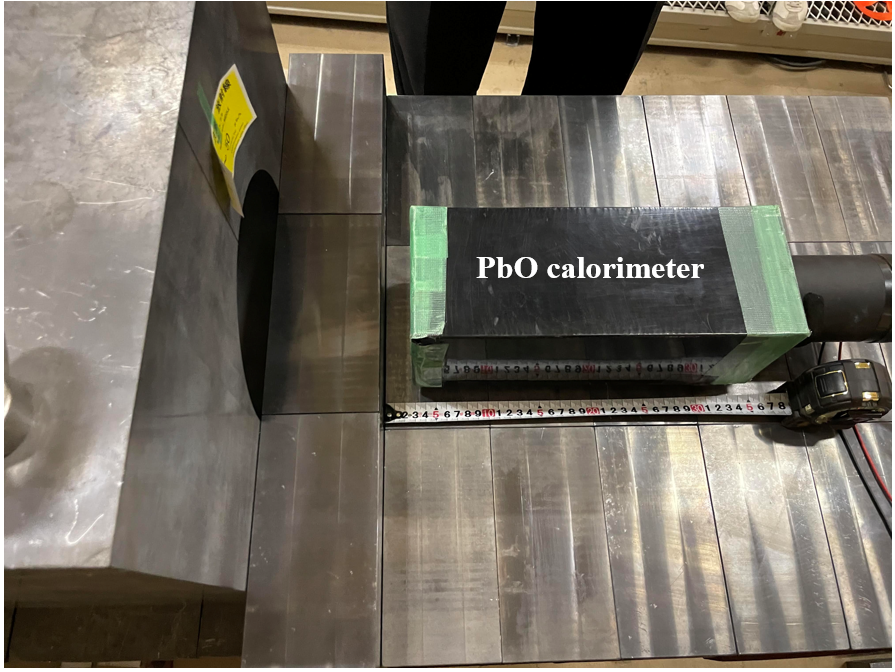}
\caption{PbO calorimeter.}
\label{fig:pbo_calorimeter}
\end{figure}

A single PbO calorimeter is used at the EBES experimental site to measure background levels and the energy of incoming particles for the search for ALPs. (Fig.~\ref{fig:pbo_calorimeter}). The PbO calorimeter was originally developed for the barrel electromagnetic calorimeter of the VENUS experiment \cite{Ogawa:1985cd}, which was conducted at the KEK TRISTAN accelerator, an electron-positron collider with a center-of-mass energy of approximately 60~GeV.

A DF6 lead-glass block is used for the PbO calorimeter, which has a cross section of $12 \times 11.6~\mathrm{cm}^2$ and a length of 30~cm, corresponding to 18 radiation lengths. The lead-glass block is glued with epoxy to a metal flange. Cherenkov light is emitted by relativistic charged particles inside the lead glass, and a 3-inch photomultiplier tube (R1911, Hamamatsu Photonics Co.) measures this light. The tube is attached to the lead-glass block through a hole in the flange. Between the photomultiplier tube and the lead-glass block, a plastic light guide with an optical filter having a cutoff wavelength of 420~nm is installed. The typical energy resolution of the calorimeter is described as $\sigma/E = 0.7\% \oplus 5.2\%/\sqrt{E}$ in Reference~\cite{Ogawa:1985cd}.

Since the detector had been left unused for several decades after the end of the VENUS experiment in 1995, its performance was studied in a testbeam, using an electron beam at the PF-AR Test Beam Line at KEK \cite{pfar}. The energy resolution of the PbO calorimeter used for the pilot run was evaluated to be $8.5\%/\sqrt{E[\mathrm{GeV}]}$ for the case in which the electron beam was injected perpendicularly at the center of the calorimeter. The details of the results are described in Appendix~\ref{sec:testbeam}.

The single PbO calorimeter is placed at the beam-center position right after the decay volume and surrounded by lead blocks with a total thickness of more than 10~cm to reduce backgrounds originating from upstream beam-loss processes.

The power supply and electronics for data taking were located in a room adjacent to the experimental area in order to avoid radiation effects. Sixty-meter-long cables were used to connect them to the PbO calorimeter. A high voltage was supplied to the PbO calorimeter by a N1470 Power Supply Module (CAEN SpA) \cite{caen_hv}. The data were collected using a 12-bit ADC (C007, HOSHIN ELECTRONICS CO., LTD.) in a CAMAC system, synchronized with the LINAC beam-injection timing signal.

\section{Data and simulation samples}
\label{chap:samples}

In this analysis, we use data collected during a dedicated EBES beam-dump pilot run with the positron beam in December 2023. The total data-taking time was approximately 2.2 hours. The incident beam consisted of positrons with an energy of $4~{\rm GeV}$ and an average bunch charge of $0.10~{\rm nC}$. The beam bunch charge was measured non-destructively using a Beam Position Monitor (BPM) installed in the LINAC beamline. When a beam bunch passes through the BPM, signals are induced on its electrodes. Since the sum of these signals is proportional to the beam bunch charge, the beam bunch charge was determined using a calibration factor established in advance. During this data-taking period, the beam repetition rate was 16.7~Hz. The number of positrons on target ($\mathrm{POT}$) was calculated as $\mathrm{POT} = (\bar{Q}_{\mathrm{bunch}}/e) \times N_{\mathrm{bunch}}$, where $\bar{Q}_{\mathrm{bunch}}$ is the average beam bunch charge measured by the BPM, $N_{\mathrm{bunch}}$ is the number of beam bunches recorded by the DAQ system, and $e$ is the elementary charge. The integrated beam exposure corresponds to $5.7\times10^{13}$ ($1.5\times10^{13}$) POT with a bias voltage of 1800 (1700) V applied to the PMT.

For the data quality assessment, we used the beam-loss monitor information described in Sec.~\ref{chap:LINAC}. Throughout the entire data-taking period, no significant increase in beam loss that could affect the analysis was observed. Therefore, the full dataset was used in the analysis.

Signal events were generated using Monte Carlo (MC) simulation. Specifically, the differential cross section for the Primakoff process given in Eq.~(\ref{eq:primakoff_dsigdt}) and the decay width for $a\to\gamma\gamma$ given in Eq.~(\ref{eq:width_agammagamma}) were implemented in the user-defined code of PHITS~\cite{Sato:2023qsv}. This implementation enables the production, transport, and decay of ALPs to be treated consistently within PHITS. The resulting event data are used as signal samples in this analysis. The signal samples were generated for multiple combinations of ALP masses and coupling constants so as to cover the sensitivity region expected in this analysis.

To efficiently generate ALP events produced via the Primakoff process, we adopted a biasing method in which the production cross section was artificially enhanced in the phase-space region relevant to the EBES acceptance. A weight was assigned to each generated event so that the correct physical probability was preserved. The final signal prediction was then obtained using the weighted event sample.

\section{Background estimation}
\label{chap:background}

The dominant background source is neutrons which are produced via photonuclear interactions of bremsstrahlung photons generated by the positron beam in the target. Instead of propagating directly to the detector, the neutrons predominantly reach the calorimeter after undergoing multiple inelastic scatterings in the shielding material.

\begin{figure}[t]
  \centering
  \includegraphics[width=0.7\textwidth]{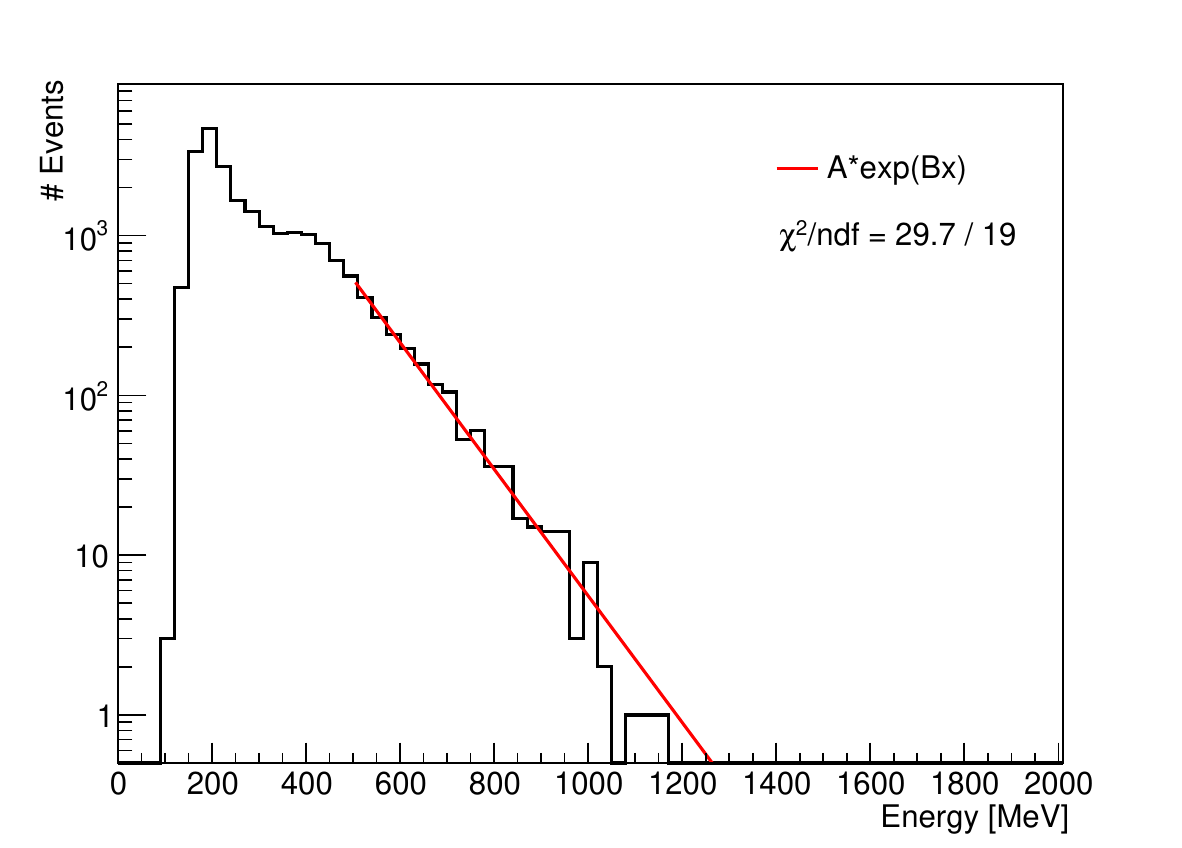}
  \caption{Energy deposit collected with a bias voltage of 1700~V applied to the PMT and scaled to correspond to an operating voltage of 1800~V. The red line shows the simple exponential function, $A \exp(Bx)$, used for the fit.}
  \label{fig:bkg}
\end{figure}

Since the simulation does not provide a sufficiently accurate description of the background, the background is estimated using a fully data-driven approach. An operating voltage of 1800~V is chosen as the optimal bias voltage applied to the Photo-Multiplier Tube (PMT), for which a large data sample is collected, while data taken at 1700~V are used for detector setup tests and background estimation, with a smaller collected data sample. The energy deposit distribution collected with a bias voltage of 1700~V and scaled to correspond to an operating voltage of 1800~V exploiting the energy resolution of the PMT, are shown in Fig.~\ref{fig:bkg}. This distribution is used as the background event distribution. 

\section{Signal event reconstruction and event selection}
\label{chap:evrec}

The energy deposit of the signal events is reconstructed using simulations by summing the energy deposited in the PbO calorimeter per unit time for each event. The total energy deposit of each event, used in the final fit, is defined as the sum of the signal and background contributions, because signal and background events can arrive simultaneously at the detector and overlap. The distribution shown in Fig.~\ref{fig:bkg} is used as the probability density function for the background energy deposit.

\begin{figure}[h]
  \centering
  \includegraphics[width=0.7\textwidth]{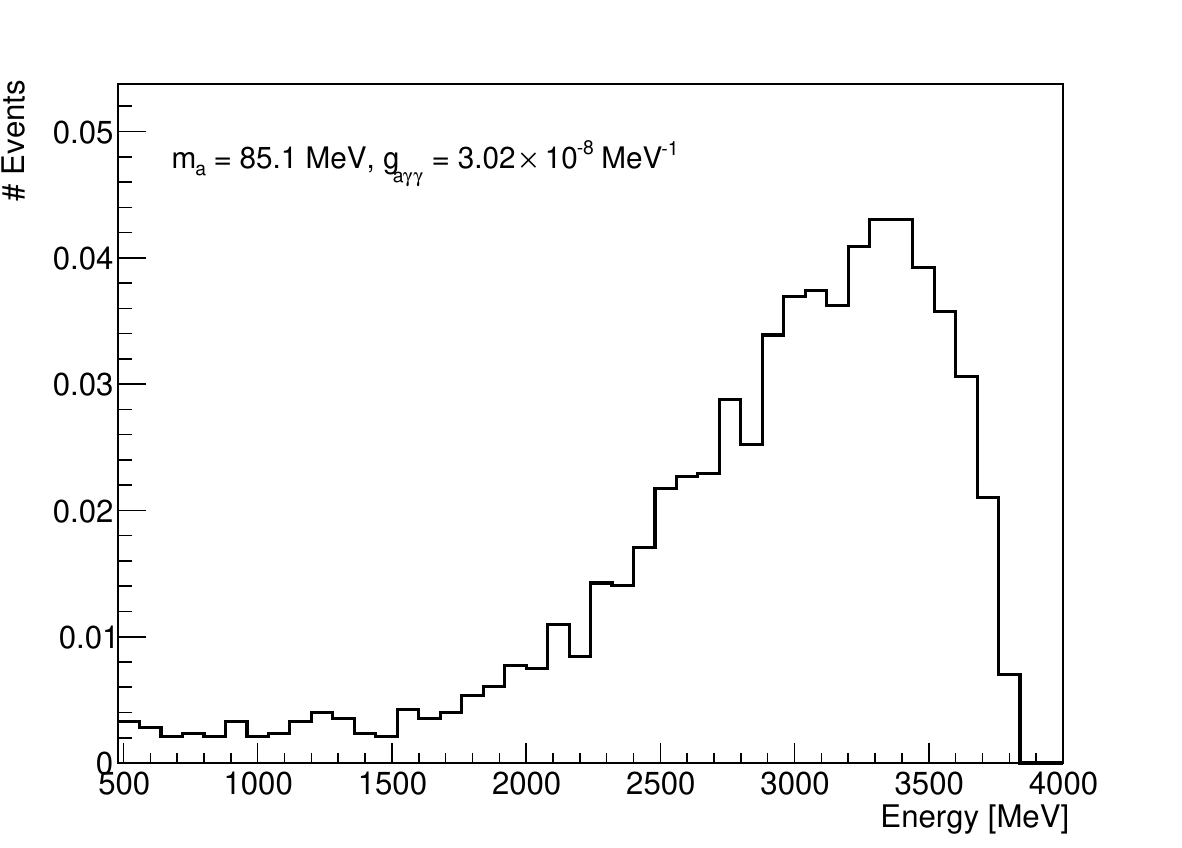}
  \caption{Energy deposit by the ALP signal events with $m_a = 85.1~\mathrm{MeV}$ and $g_{a\gamma\gamma} = 3.02 \times 10^{-8}$.}
  \label{fig:signal}
\end{figure}

An example of the simulated energy deposit by the ALPs signal events is shown in Fig.~\ref{fig:signal}. A signal point close to the highest mass limit obtained in this analysis is shown as a representative example. The background contribution is added as described above, and the number of events are normalized to the amount of collected data. The energy deposit by the signal events is determined by the interplay between the production cross section, the decay length which is determined by the decay width given in Eq.~(\ref{eq:width_agammagamma}), governing where the particles decay, and the Lorentz boost of the ALPs. As the mass and the coupling increase, the decay width becomes larger, leading to a shorter decay length. In this regime, a substantial fraction of the particles decay in the shielding material located upstream of the detector. As a consequence, only highly boosted particles reach the detector, resulting in a hard energy-deposit spectrum.

Compared with the background events shown in Fig.~\ref{fig:bkg}, the signature of the ALPs signal events is a large energy deposit extending up to 4000~MeV. The background distribution in Fig.~\ref{fig:bkg} is fitted in the range $500 < E_{\rm dep} < 1500~\mathrm{MeV}$ with a simple exponential function, $A \exp(Bx)$. Signal events are selected requiring an energy deposit above the signal-region threshold. The threshold to define signal region is then determined as the energy at which the expected number of background events falls below 0.1, resulting in an energy deposit greater than 1500~MeV.

To avoid potential analysis bias, the signal region of the data collected with the PMTs operated at 1800~V was blinded until the analysis strategy was finalized. The signal region was unblinded only after all analysis procedures had been fixed.

\section{Systematic uncertainties}
\label{chap:systematics}

POT normalization, threshold difference in the energy deposit in the calorimeter, and energy resolution of the PbO calorimeter on the number of signal events are considered as sources of the systematic uncertainty in the analysis. The statistical uncertainty from the limited number of MC simulated signal events is also taken into account.

The systematic uncertainty associated with the POT normalization is evaluated conservatively using cosmic-muon data and 7~GeV beam data. Although the absolute value of the beam bunch charge is monitored by the LINAC instrumentation, an additional data-driven evaluation is adopted under the pilot run conditions. For the 7~GeV beam, the energy deposited in the PbO calorimeter is dominated by muons produced by bremsstrahlung photons. 
To relate the muon-induced energy deposit to the beam bunch charge, a MC simulation that has been well validated for a similar experimental setup~\cite{Sakaki:2020cux} is employed and used to evaluate both the energy flux of bremsstrahlung-induced muons and the corresponding energy deposit in the calorimeter. A scale factor accounting for the difference between data and simulation for single-muon response is derived from cosmic-muon data and applied to the 7~GeV beam simulation. The ratio between the corrected simulation and the corresponding 7~GeV beam data is then taken as the systematic uncertainty associated with the POT. This treatment also covers residual differences between data and simulation, and results in a 20\% impact on the final limit. In a future upgrade, an additional device is planned to be installed in order to reduce this uncertainty.

The systematic uncertainty associated with the threshold difference in the energy deposit in the calorimeter is evaluated by varying the threshold to 1200~MeV and 1800~MeV, instead of the nominal 1500~MeV, and examining the resulting change in the number of signal events in the signal region.
The number of signal events in the signal region varies by approximately 10--20\%, depending on the mass and the coupling. This results in an impact of approximately 5\% on the final limit.

The energy resolution of the PbO calorimeter is evaluated to be $8.5\%/\sqrt{E[\mathrm{GeV}]}$ as described in Sec. \ref{sec:pbo}. The reconstructed signal energy deposits are varied within the detector energy resolution, and the resulting change in the number of signal events passing the energy threshold of the signal region is taken as the systematic uncertainty on the signal yield. The gain stability of the PMT in the PbO calorimeter was measured to be better than 0.5\% in the testbeam under similar operation conditions and over a comparable operating period. This uncertainty is significantly smaller than that arising from the energy resolution and is therefore neglected in the evaluation of the systematic uncertainty.

\section{Results}
\label{chap:results}

\begin{figure}[t]
  \centering
  \includegraphics[width=0.7\textwidth]{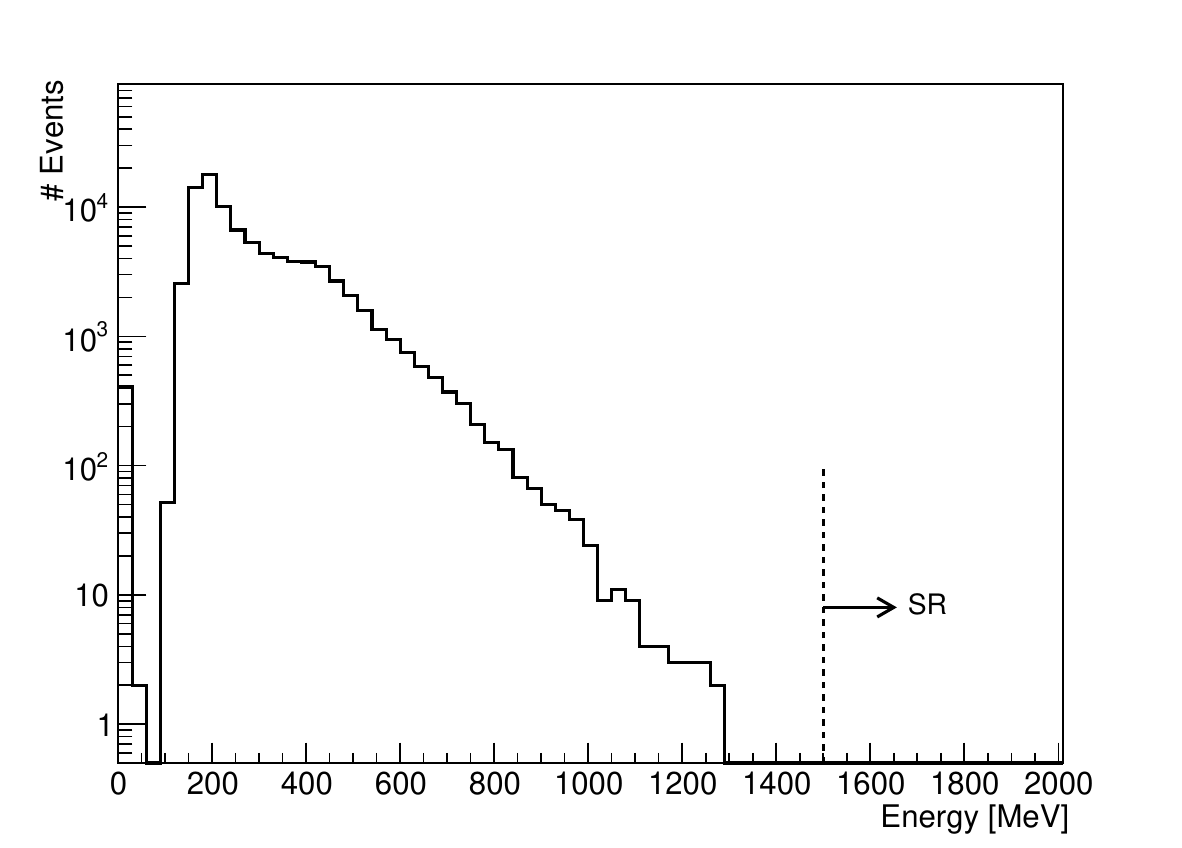}
  \caption{Energy deposit collected with a bias voltage of 1800~V applied to the PMT. The signal region is defined as the region with energy deposits above 1500 MeV, as indicated by the arrow.}
  \label{fig:data1800V}
\end{figure}

The signal region is designed to have a negligible expected background contribution. Fig.~\ref{fig:data1800V} shows the energy deposit collected with a bias voltage of 1800 V applied to the PMT. Since no event is observed in the signal region, 90\% confidence level upper limit is set on the ALP mass-coupling plane.

The statistical analysis is performed using TRExFitter~\cite{TRExFitter}, based on the event yield in the signal region, without exploiting the shape information of the energy distribution. Upper limits on the signal strength are derived using a frequentist toy MC approach based on a profile likelihood ratio test statistic. The signal strength, denoted by $\mu$, is treated as the parameter of interest, while the systematic uncertainties described in Sec.~\ref{chap:systematics} are included as nuisance parameters. For each mass--coupling point, pseudo-experiments are generated for each scanned value of $\mu$, and the 90\% confidence level upper limit is determined from the resulting test statistic distributions.

\begin{figure}[t]
  \centering
  \includegraphics[width=0.7\textwidth]{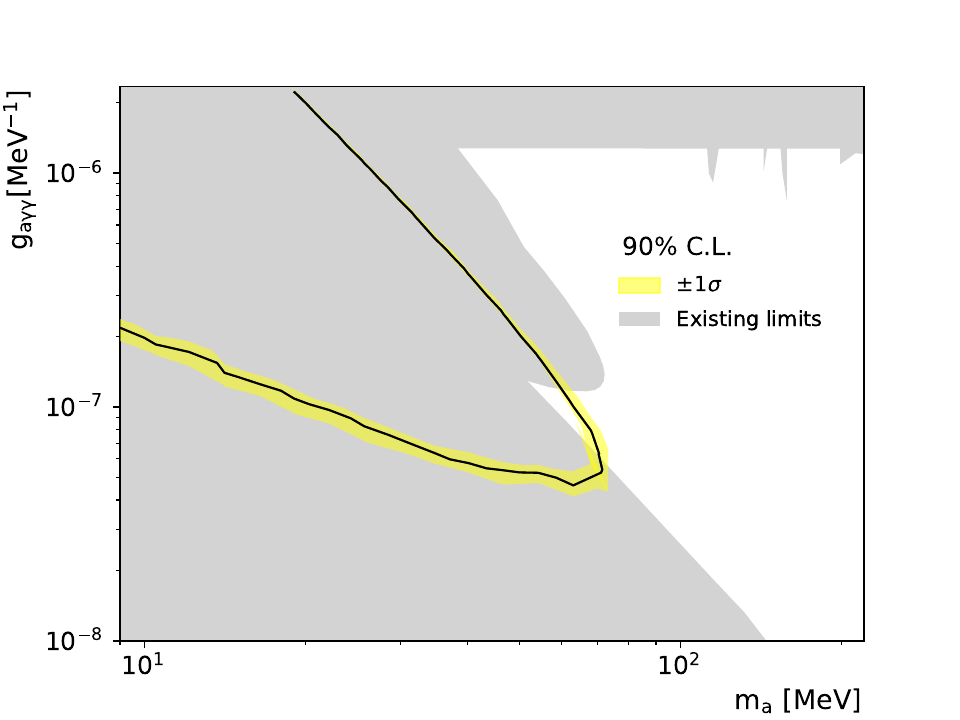}
  \caption{The 90\% confidence level upper limits in the ALP mass--coupling plane. The yellow band indicates the $\pm1\sigma$ uncertainty, including both statistical and systematic uncertainties.
}
  \label{fig:contour}
\end{figure}

Figure~\ref{fig:contour} shows the 90\% confidence level upper limits in the ALP mass--coupling plane. 
The gray shaded region indicates parameter space excluded by previous experiments, including E137~\cite{Bjorken:1988as}, 
NuCal~\cite{Blumlein:1990ay,Dobrich:2019dxc}, OPAL~\cite{OPAL:2002vhf,Knapen:2016moh}, 
PrimEx~\cite{PrimEx:2010fvg,Aloni:2019ruo}, NA64~\cite{NA64:2020qwq}, Belle~II~\cite{Belle-II:2020jti}, 
and FASER~\cite{FASER:2024bbl}. 
The results of the EBES pilot run extend the experimental coverage into a region of parameter space not explored by the previous searches.

\section{Conclusions}
The results of a search for ALPs using the pilot run data of the EBES experiment are presented. The analysis was performed using data taken with a 4~GeV positron beam at the KEK LINAC in December 2023. A signal region was defined in the high-energy tail of the PbO calorimeter response, where the expected background contribution is negligible. After unblinding, no events were observed in the signal region.

Upper limits at the 90\% confidence level were derived in the ALP mass--coupling plane. The obtained limits extend the experimental coverage into a region of parameter space not explored by previous searches, as shown in Fig.~\ref{fig:contour}. This result demonstrates that the EBES setup has sufficient sensitivity to perform a competitive search for sub-GeV ALPs even with the pilot run configuration.

The pilot run also provided important information on the background level at the experimental site and validated the analysis procedure used in this study. Since the full EBES experiment started physics data-taking in the autumn of 2025 with an upgraded detector configuration, a substantially broader region of parameter space will be explored in future analyses.

\section*{Acknowledgments}
This work was supported by JSPS KAKENHI Grant Numbers JP21H05466, JP23K25885, JP24K07077, and JP25K01009. We are grateful to the KEK LINAC group for their support in realizing the EBES experiment. We also thank Kazuro Furukawa and Mitsuhiro Yoshida for their valuable advice and support in the early development of the experiment. 
We gratefully acknowledge the support and assistance of all personnel involved in the PF-AR Test Beam Line of the Instrumentation Technology Development Center at KEK.
We acknowledge valuable technical support from the KEK Radiation Science Center and the KEK Mechanical Engineering Center in the preparation of the shielding and beam-dump system, and thank Kazuhiko Iijima and Toshikazu Takatomi for their helpful support. We also thank Futaba Kogyo Co., Ltd. for valuable advice in the construction of the early experimental setup.

\appendix
\renewcommand{\thefigure}{A\arabic{figure}}
\setcounter{figure}{0}

\begin{figure}[t]
\centering
\begin{minipage}[b]{0.48\textwidth}
    \centering
    \includegraphics[width=\textwidth]{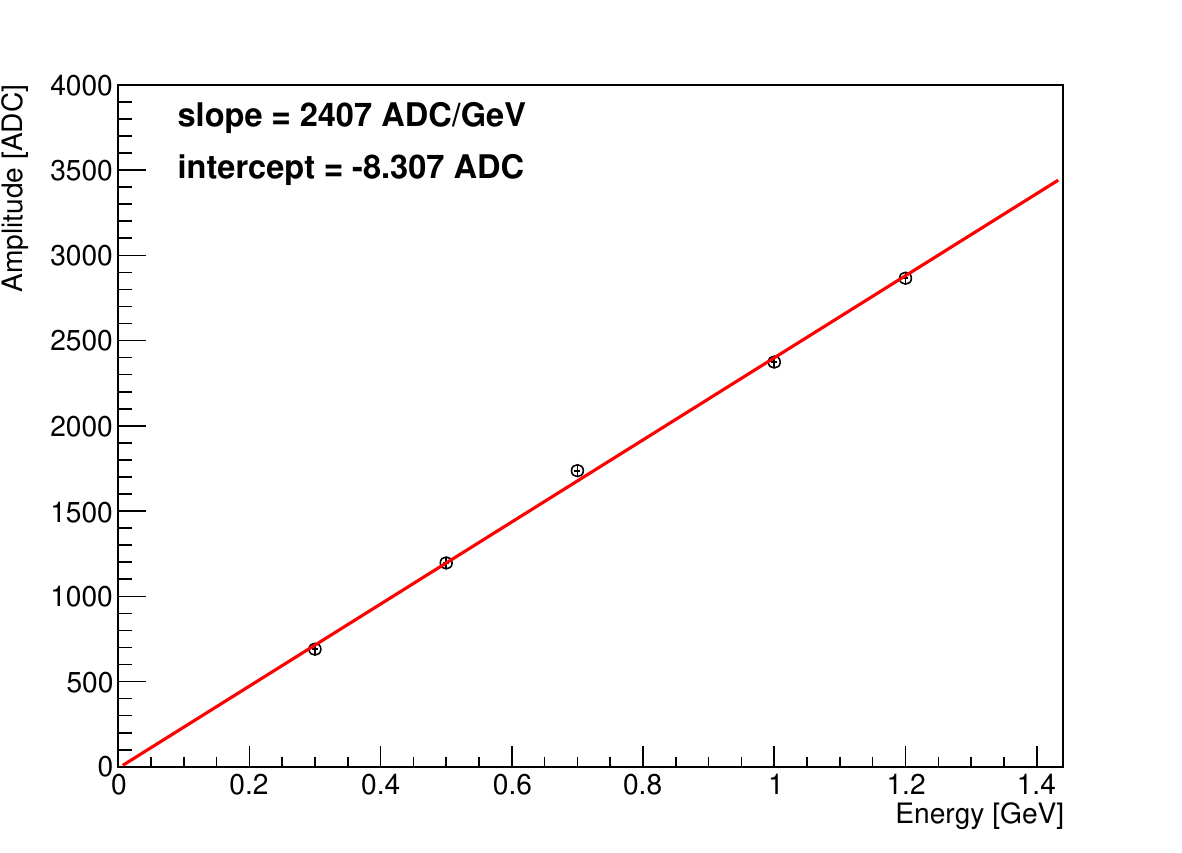}
    
    {\small (a) Linearity}
\end{minipage}
\begin{minipage}[b]{0.48\textwidth}
    \centering
    \includegraphics[width=\textwidth]{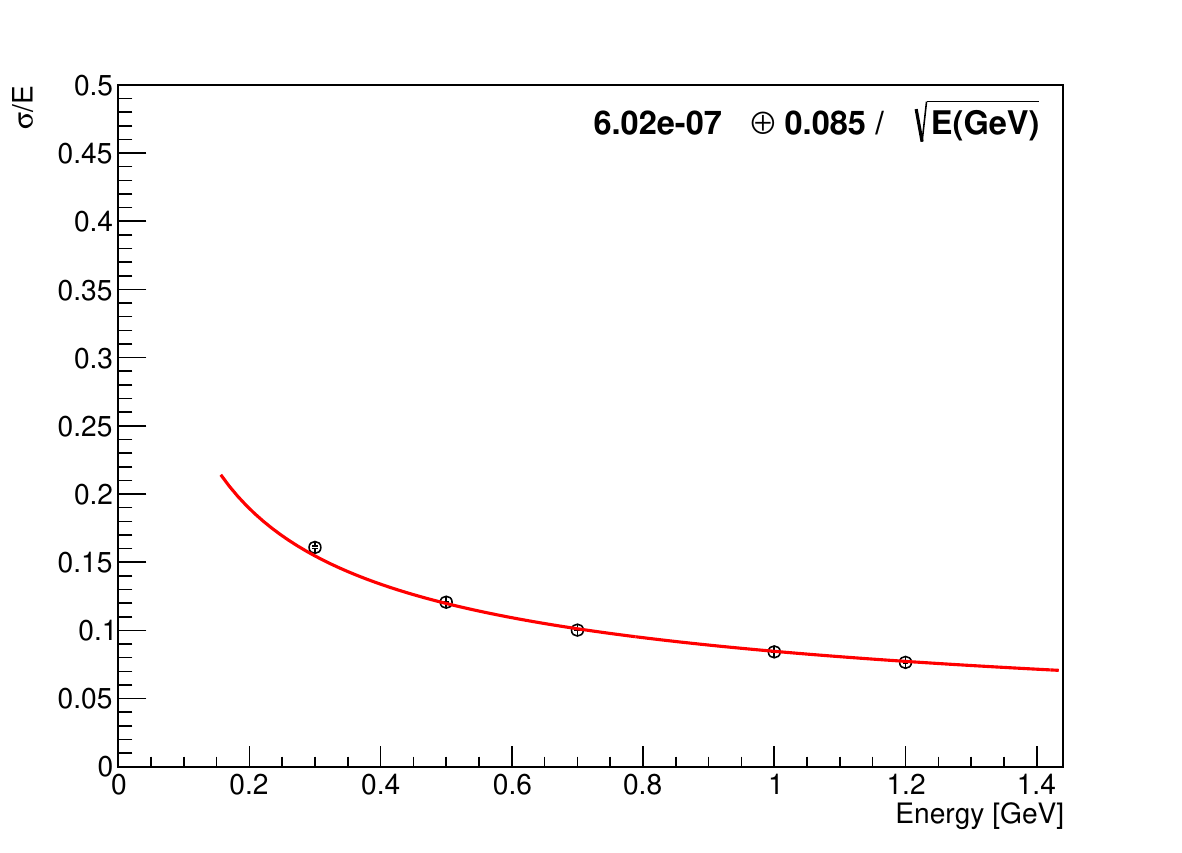}
    
    {\small (b) Resolution}
\end{minipage}
\begin{minipage}[b]{0.48\textwidth}
    \centering
    \includegraphics[width=\textwidth]{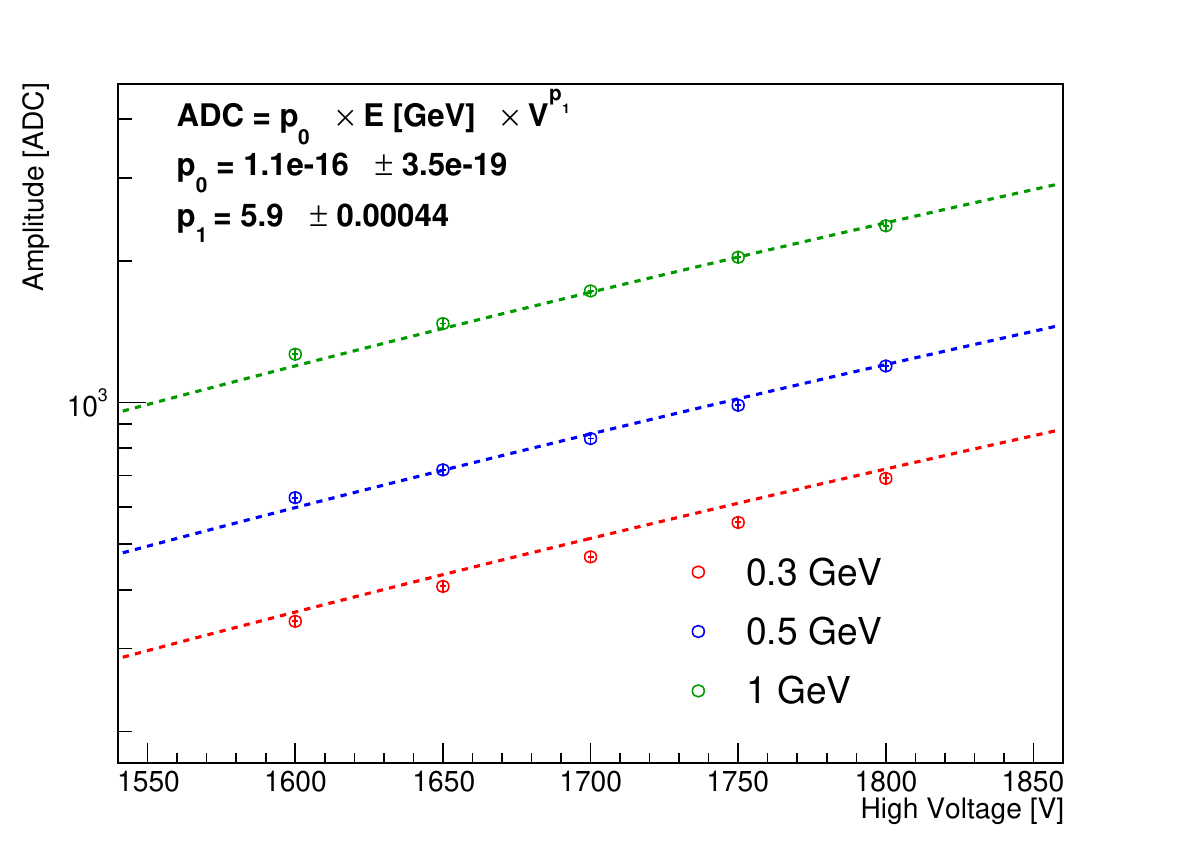}
    
    {\small (c) HV dependence}
\end{minipage}
\caption{The linearity (a), the energy resolution (b), and the HV dependence (c) of the PbO calorimeter used for the pilot run. The PMT operating voltage was set to 1800~V for (a) and (b).}
\label{fig:pbo_resolution}
\end{figure}

\section{Testbeam for PbO calorimeter} \label{sec:testbeam}
The PbO calorimeter used in this study was calibrated at the KEK PF-AR test-beam line. The PF-AR facility provides an electron test beam derived from the 6.5 GeV PF-AR storage ring. In this beamline, a tungsten wire is inserted into the circulating beam to generate secondary electrons, which are then extracted and transported to the experimental area. Electron beams with energies up to 5 GeV are available, which sufficiently cover the energy range assumed for the ALP search. 

The calibration measurements were performed during 5–10 June 2024. A trigger was formed by a plastic scintillation counter installed upstream of the PbO calorimeter along the beam axis. For each triggered event, the signal from the photomultiplier tube (PMT) coupled to the PbO crystal was digitized using the CAMAC ADC used in the main experiment. Multiple runs were taken to characterize (i) the relation between the beam energy and ADC count, its linearity and the energy resolution of the detector, (ii) the dependence on the PMT high voltage (HV), and (iii) module-to-module variations in gain and resolution.

Figure~\ref{fig:pbo_resolution} summarizes the linearity, the energy resolution, and the HV dependence of the PbO calorimeter used in the main experiment. A clear linear correlation was observed between the incident beam energy and the ADC response over the explored energy range. This response was used to establish the energy conversion from ADC counts to deposited energy for the main experiment. The absolute scale of the energy of the beam line may have uncertainty of up to 10\%, which is the dominant uncertainty to the ADC-energy conversion. The HV dependence is used for the correction of Fig.~\ref{fig:bkg}.

The energy resolution was parameterized by a stochastic term of
\begin{equation}
\frac{\sigma_E}{E} = \frac{8.5\%}{\sqrt{E[\mathrm{GeV}]}} ,
\end{equation}
as extracted from the signal distribution width. 
Since the constant term of the energy resolution is consistent with zero, as shown in Fig.~\ref{fig:pbo_resolution}(b), we expect the energy spread of the beam to be negligibly small, which is a conservative assumption for the systematic uncertainty.
This performance satisfies the requirements of the present experiment and provides sufficient precision for the ALP search.

\bibliographystyle{ptephy}
\bibliography{references}

\begin{thebibliography}{10}

\bibitem{Peccei:1977hh}
R.~D. Peccei and Helen~R. Quinn, Phys. Rev. Lett., {\bf 38}, 1440--1443 (1977).

\bibitem{Weinberg:1977ma}
Steven Weinberg, Phys. Rev. Lett., {\bf 40}, 223--226 (1978).

\bibitem{Wilczek:1977pj}
Frank Wilczek, Phys. Rev. Lett., {\bf 40}, 279--282 (1978).

\bibitem{Svrcek:2006yi}
Peter Svrcek and Edward Witten, JHEP, {\bf 06}, 051 (2006),
  {{hep-th/0605206}}.

\bibitem{Arvanitaki:2009fg}
Asimina Arvanitaki, Savas Dimopoulos, Sergei Dubovsky, Nemanja Kaloper, and
  John March-Russell, Phys. Rev. D, {\bf 81}, 123530 (2010),
  {{arXiv:0905.4720}}.

\bibitem{Beacham:2019nyx}
J.~Beacham et~al., J. Phys. G, {\bf 47}(1), 010501 (2020),
  {{arXiv:1901.09966}}.

\bibitem{Antel:2023hkf}
C.~Antel et~al., Eur. Phys. J. C, {\bf 83}(12), 1122 (2023),
  {{arXiv:2305.01715}}.

\bibitem{Ishikawa:2021qna}
Akimasa Ishikawa, Yasuhito Sakaki, and Yosuke Takubo, PTEP, {\bf 2022}(11),
  113B05 (2022),  {{arXiv:2107.06431}}.

\bibitem{Akemoto:2013lxa}
Mitsuo Akemoto et~al., PTEP, {\bf 2013}, 03A002 (2013).

\bibitem{Furukawa:2022wzc}
K.~Furukawa et~al., JACoW, {\bf IPAC2022}, THPOST012 (2022),
  {{arXiv:2207.01192}}.

\bibitem{Tsai:1986tx}
Yung-Su Tsai, Phys. Rev. D, {\bf 34}, 1326 (1986).

\bibitem{Dusaev:2020gxi}
R.~R. Dusaev, D.~V. Kirpichnikov, and M.~M. Kirsanov, Phys. Rev. D, {\bf
  102}(5), 055018 (2020),  {{arXiv:2004.04469}}.

\bibitem{Ogawa:1985cd}
K.~Ogawa, K.~Hayashi, T.~Sumiyoshi, F.~Takasaki, Y.~Teramoto, T.~Uehara,
  S.~Sugimoto, H.~Kusumoto, J.~Iwahori, and H.~Yoshida, Nucl. Instrum. Meth. A,
  {\bf 243}, 58--66 (1986).

\bibitem{pfar}
{KEK},
\newblock Test beam line at kek pf-ar,
\newblock \url{https://itdc.kek.jp/en/testBeamLine/index.html} (2026).

\bibitem{caen_hv}
{CAEN},
\newblock N1407 power supply modules,
\newblock \url{https://www.caen.it/products/n1470/} (2026).

\bibitem{Sato:2023qsv}
Tatsuhiko Sato et~al., J. Nucl. Sci. Tech., {\bf 61}(1), 127--135 (2024).

\bibitem{Sakaki:2020cux}
Yasuhito Sakaki, Yoshihito Namito, Toshiya Sanami, Hiroshi Iwase, and Hideo
  Hirayama, Nucl. Instrum. Meth. A, {\bf 977}, 164323 (2020),
  {{arXiv:2004.00212}}.

\bibitem{TRExFitter}
{ATLAS Collaboration},
\newblock {TRExFitter, Zenodo},
\newblock https://doi.org/10.5281/zenodo.14845712 (2026).

\bibitem{Bjorken:1988as}
J.~D. Bjorken, S.~Ecklund, W.~R. Nelson, A.~Abashian, C.~Church, B.~Lu, L.~W.
  Mo, T.~A. Nunamaker, and P.~Rassmann, Phys. Rev. D, {\bf 38}, 3375 (1988).

\bibitem{Blumlein:1990ay}
J.~Blumlein et~al., Z. Phys. C, {\bf 51}, 341--350 (1991).

\bibitem{Dobrich:2019dxc}
Babette D{\"o}brich, Joerg Jaeckel, and Tommaso Spadaro, JHEP, {\bf 05}, 213,
  [Erratum: JHEP 10, 046 (2020)] (2019),  {{arXiv:1904.02091}}.

\bibitem{OPAL:2002vhf}
G.~Abbiendi et~al., Eur. Phys. J. C, {\bf 26}, 331--344 (2003),
  {{hep-ex/0210016}}.

\bibitem{Knapen:2016moh}
Simon Knapen, Tongyan Lin, Hou~Keong Lou, and Tom Melia, Phys. Rev. Lett., {\bf
  118}(17), 171801 (2017),  {{arXiv:1607.06083}}.

\bibitem{PrimEx:2010fvg}
I.~Larin et~al., Phys. Rev. Lett., {\bf 106}, 162303 (2011),
  {{arXiv:1009.1681}}.

\bibitem{Aloni:2019ruo}
Daniel Aloni, Cristiano Fanelli, Yotam Soreq, and Mike Williams, Phys. Rev.
  Lett., {\bf 123}(7), 071801 (2019),  {{arXiv:1903.03586}}.

\bibitem{NA64:2020qwq}
D.~Banerjee et~al., Phys. Rev. Lett., {\bf 125}(8), 081801 (2020),
  {{arXiv:2005.02710}}.

\bibitem{Belle-II:2020jti}
F.~Abudin{\'e}n et~al., Phys. Rev. Lett., {\bf 125}(16), 161806 (2020),
  {{arXiv:2007.13071}}.

\bibitem{FASER:2024bbl}
Roshan Mammen~Abraham et~al., JHEP, {\bf 01}, 199 (2025),
  {{arXiv:2410.10363}}.

\end{thebibliography}
%

\vspace{0.2cm}
\noindent


\let\doi\relax


\end{document}